\documentclass[a4paper, amsfonts, amssymb, amsmath, reprint, showpacs, showkeys, nofootinbib,prl, twoside, floatfix]{revtex4-2}
\usepackage[english]{babel}
\usepackage[utf8]{inputenc}
\usepackage{upgreek}
\usepackage[colorinlistoftodos, color=green!40, prependcaption]{todonotes}
\usepackage{amsthm}
\usepackage{mathtools}
\usepackage{siunitx}
\usepackage{xcolor}
\usepackage{graphicx}
\usepackage[autostyle]{csquotes}
\usepackage[left=16mm,right=16mm,top=35mm,bottom=20mm,columnsep=14pt]{geometry} 
\usepackage{adjustbox}
\usepackage{placeins}
\usepackage{multirow}
\usepackage[T1]{fontenc}
\usepackage{lipsum}
\usepackage{url}
\usepackage{natbib,hyperref}
\hypersetup{colorlinks=true, urlcolor=blue, linkcolor=blue, citecolor=blue}
\usepackage{multirow, booktabs}
\usepackage{epstopdf}

\begin{document}
\renewcommand{\figurename}{Fig.{}}
\title{Observation of rovibrationally coupled bi-modality and speed-dependent orientation in DEA dynamics of OCS: reveals partial correlations among point group symmetries }

\author{Narayan Kundu\textit{$^{1}$}}
\author{Vikrant Kumar\textit{$^{1}$}}
\author{Tejas Jani\textit{$^{2}$}}
\author{Minaxi Vinodkumar\textit{$^{2}$}}
\author{Dhananjay Nandi\textit{$^{1,3}$}}%
\email{dhananjay@iiserkol.ac.in}
\affiliation{$^1$Indian Institute of Science Education and Research Kolkata, Mohanpur-741246, India.\\ $^2$VP and RPTP Science College, Vallabh Vidyanagar, Mota Bazar, Gujrat$-$388120, India.\\ $^3$Center for Atomic, Molecular and Optical Sciences $\&$ Technologies, Joint initiative of IIT Tirupati $\&$ IISER Tirupati, Yerpedu, 517619, Andhra Pradesh, India}





\begin{abstract}
Dissociative electron attachment (DEA) to gas-phase carbonyl sulfide (OCS) has been studied diligently using the time-of-flight (TOF) based state-of-the-art velocity map imaging (VMI) technique. Three well-resolved DEA resonances are observed at 5.0, 6.5 and 10.0 eV incident electron energies along with a weak structure at 8.0 eV. The velocity slice images (VSI), Kinetic energy (KE) and angular distributions (AD) for the fragmented S$^-$ anions are obtained using the wedge slicing technique \cite{moradmand2013_wedge,nag2019dissociative,kundu2021effect}. The KE distributions for the S$^-$ nascent fragments reveal bi-modality with rovibrational signatures. The ADs substantiate speed-dependent angular anisotropy demand the existence of partial correlations among three different point group symmetries (C$_{S}$, C$_{2V}$ \& C$_{\infty V}$), confirmed through an in-plane bending mode of vibration with the axial recoil breakdown. Theoretical calculations using R-matrix and density functional approaches strongly support the experimental observations. 

\end{abstract}

\maketitle

Dynamics study using charged fragments has marked a new realm in the fields of electron-induced chemistry, plasma physics, radiation therapy, astrochemistry, semiconductor fabrication and surface sciences \cite{coburn1979ion,boamah2014low,krishnakumar2018symmetry,charged_plasma_accelerator,charged_plasma}. Describing the existence of electronic resonances in low-energy electron molecule collisions through dissociative electron attachment (DEA) are proving of fundamental importance in recent times \cite{krishnakumar2018symmetry,fabrikant2017recent,dea_hnco_prl,dea_h2d2_prl}, ranging from DNA damage \cite{li2_dna_damage_jacs,nikjoo1997computational,bao_dna_damage_pnas,pan_dna_prl,dea_uracil_prl} to molecular cluster formation \cite{bowen_cluster,illenberger_chem_rev}. In DEA, resonances occur at some specific beam energies whereon the colliding particles generate quasi-bound temporary negative ion (TNI) complexes \cite{whelan2006electron,krishnakumar1998dissociative}; resulting in rapid variations in the energy dependence of the inelastic scattering cross-sections. Two different modes of the resonantly formed TNI  state, one through nuclear motion (dissociation) and the other by electronic motion (autodetachment), provide us with an ideal ground to study the coupling between electronic and nuclear degrees of freedom \cite{krishnakumar1998dissociative}. However, the characteristics of these negative ion-resonances are mainly governed by the electron correlation phenomenon.

We have investigated the DEA dynamics of naturally present most abundant sulfur-containing gaseous linear triatomic OCS \cite{hattori2020constraining}, observed in the terrestrial, as well as extraterrestrial environments like the Venusian atmosphere \cite{hong1997formation,ueno_young_sun_para,charnley1997sulfuretted,mumma2003remote,kamp1990radiative,fer_ocs_observe,woodney1997detection,charnley2004observational,van2014astrochemistry} that acts as a coupling catalyst for production of peptides from amino acids \cite{leman_ocs_science}. In this letter, we reveal the cross-sectionally highest anionic channel reporting time-gated wedge-shaped VSIs, KE \& ADs; manifesting kinematically complete molecular dynamic pieces of information:
\begin{align}
    \rm{OCS} + e^- \rightarrow (\mbox{OCS}^-)^* \ & \longrightarrow \mbox{S}^- + \mbox{CO}
\end{align}

\begin{figure}[htb]
    \centering
    \includegraphics[scale=0.41]{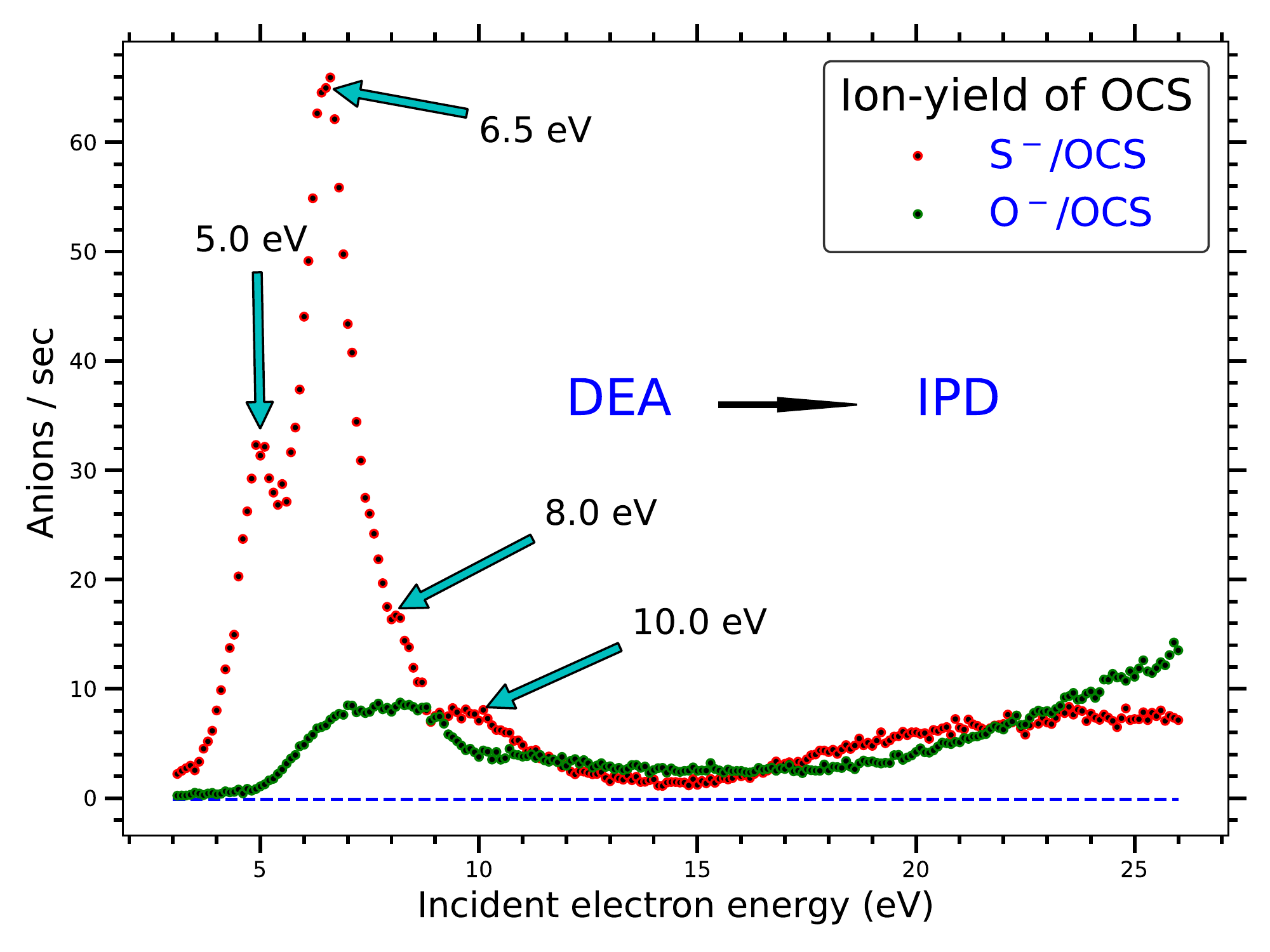}
    \caption{Excitation function (ion-yield) of S$^-$/OCS and O$^-$/OCS anions. Figure depicts ion-pair dissociation thresholds are different for both the anions.}
    \label{fig:ion_yield}
\end{figure}
Excitation function of S$^-$/OCS as displayed in Fig. \ref{fig:ion_yield} clearly resolves three different DEA resonances positioned at 5.0, 6.5 and 10.0 eV incident electron energies, agrees well with others reports \cite{iga1995,iga1996_elsever,abouaf1976,dillard1968ion,tronc1982zero,macneil1969negative,ziesel1975s,hubin1976dissociative}. A weak resonance structure observed near the 8.0 eV incident electron energy is not disclosed in previous reports \cite{iga1995,abouaf1976,dillard1968ion,tronc1982zero,macneil1969negative,ziesel1975s,hubin1976dissociative,iga1996_elsever}. Interestingly, we observe two different ion-pair generation threshold for the fragmented S$^-$ \& O$^-$ anions that strongly demands the existence of channel specific ion-pair threshold, as reflected in Fig. \ref{fig:ion_yield}.

\begin{figure*}
    \centering
    \includegraphics[scale=0.85]{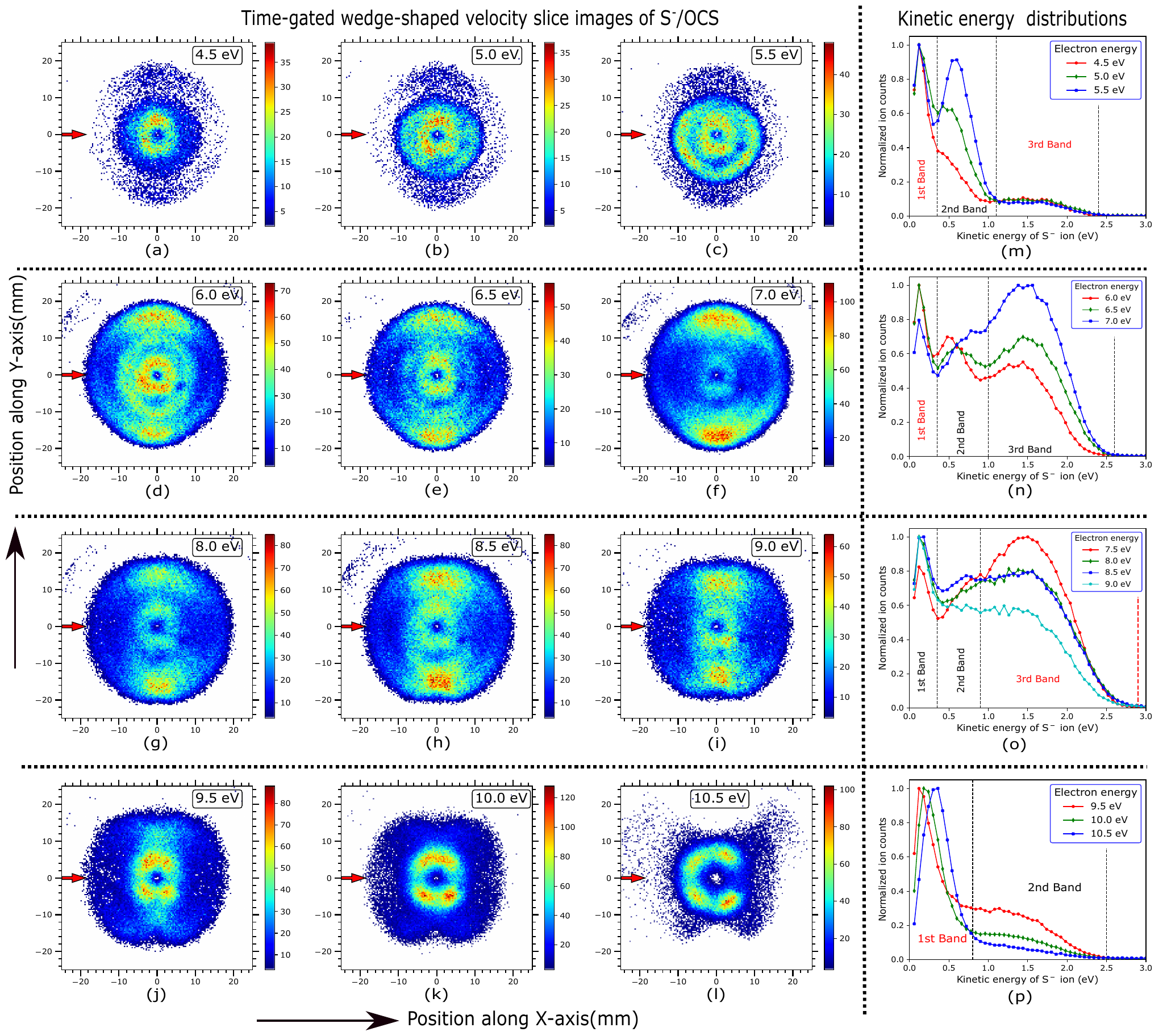}
    \caption{(a-i) 50 ns time-gated wedge-shaped velocity slice images are shown as a function of electron beam energy. For each VSI, the beam energy is written within a box at the upper right corner. The red arrow used from left to right signifies the direction of the electron beam. (m-p) KE distributions of S$^-$/OCS for four distinct DEA resonances.}
    \label{fig:all_vsi}
\end{figure*}
Fig. \ref{fig:all_vsi} represents time-gated wedge-shaped VSIs of S$^-$/OCS as a function of electron beam energy. A close look at the VSIs unfolds four distinct slice images with three dissociative KE bands (tri-modal) at 5.0, 6.5, 8.0 eV resonances, while the 10.0 eV resonance exhibits only two. This tri-modality in KE distribution signifies three distinct dissociative pathways for the TNI of OCS, well behaved with the KE distributions as insipidly outlined by Li \textit{et al.} \cite{li2016s} using the time-gated parallel slicing technique. An unexpected intensity boost in the highest momentum band is observed at 6.0 eV, attaining a maximum at 7.0 eV beam energy, rationalised herein. 

Generally, vibrational spectra of di-atomic molecules result from the change in dipole moment vector along the molecular axis, wherein stretching vibration followed by bond length variations play a crucial role in controlling the dynamics. The situation changes radically for a tri-atomic linear molecule, exhibiting four irreducible vibrational modes (two non-degenerate stretching modes associated with $\Sigma$ symmetry \& two degenerate bending modes associated with $\Pi$ symmetry). Here, in and out of phase bending mode vibrations is permitted through bond angle variations, conserving total angular momentum that allows transition with the same rotational quantum numbers in the ground and excited vibrational states for all the populated rotational levels. Such vibronic coupling in an open shell system imparts a non-zero vibrational component of angular momentum projected along the molecular axis \cite{kellman_pra}, a rationale for perceiving intensity maximisation of the high momentum anionic band perpendicular to the beam direction. Breakdown of Born-Oppenheimer approximation occurs when two-fold $\Pi$ state is annihilated by the bending vibration, occurring due to the Renner-Teller effect (RTE) \cite{peric_rt_effect,brown_rt_hamiltonian,herzberg1933teller,renner1934theorie,herzberg2013molecular_rt,bunker2006molecular_rt,Herzberg_three,1966msms.book.....H,jungen2019renner,wormer2003original,cam_b3lyp_rt_effect}. However, vibronic coupling generates an angular momentum that splits rotational levels, mainly through axial rotation. Hence, the dynamics of triatomic molecules strongly insists on rovibronic interpretation \cite{zak2017ro_vibronic}. Nevertheless, the neutral di-atomic counterpart (CO) can be fragmented with such highly populated rotational states, which arises due to rovibronically coupled strong bending forces in triatomic OCS \cite{wei2016photodissociation}. A few reports on photo-dissociation and DEA to OCS \cite{sato1995ion,sugita2000effect,sivakumar1988state,bai2017photo_jcp,bai2017photo_chainese,mcbane2013ultraviolet,suzuki1998nonadiabatic,iga1995,iga1996_elsever} reported that CO fragments are dominantly associated with highly excited rotational levels in the vibrational ground state, which confirms that CO stretching acts as a spectator in the dynamics. Our observations agree with such reports for rovibronically coupled high momentum bands. Using energy-momentum conservation, we can say that anions with a lower KE band correspond to higher excited states of the neutral fragment. Thus, for low-speed S$^-$ ions i.e. the first \& second KE bands in first two resonances, CO can be fragmented into its rovibrationally excite ($\nu \neq 0$, J) electronically ground states \cite{iga1995,iga1996_elsever}, followed by non-adiabatic bending dissociation \cite{suzuki1998nonadiabatic}. Recently, using 2+1 Resonance-enhanced multi-photon ionisation (REMPI) to OCS photo-dissociation, Gunthardt \textit{et al.} \cite{gunthardt2019anomalous} studied anomalous intensity spectrum for the CO photo-fragments and assign rovibrationally excited states to describe distinct KE bands of CO, agreements well with our report.
\begin{figure*}
    \centering
    \includegraphics[scale=0.75]{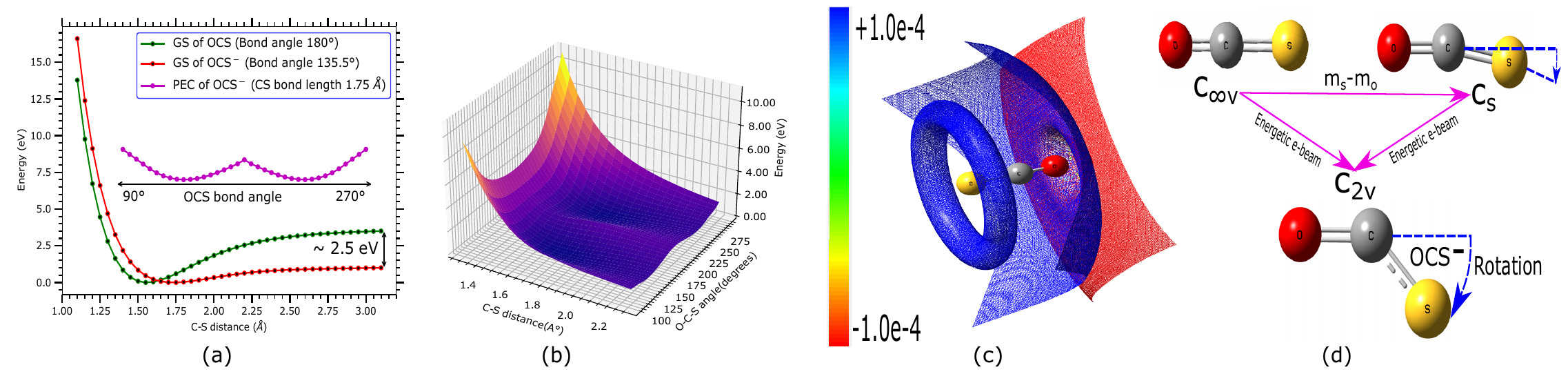}
    \caption{(a) The plot shows GS PEC for neutral OCS (green color) and anionic OCS (red color). (b) the GS potential energy surface (PES) of OCS$^-$. (c) The electrostatic potential (ESP) map 0.0004 isovalue surface from total self consistent field density for neutral OCS indicates a higher electron density in the vicinity of oxygen than the sulfur atom that signifies electron absorbing DEA cross-section is more for S$^-$ fragments as reflected in Fig. \ref{fig:ion_yield}. (d) Electron attachment provides bent geometry for stable anionic OCS with C$_{2v}$ point group. }
    \label{fig:dft}
\end{figure*}
To confirm the hypothesis of bending vibrations, an \textit{ab initio} density functional theory (DFT) calculation has been carried out using Gaussian 09 package \cite{gaussian_09} to probe the ground state (GS) of neutral and anionic OCS. We have visualised the GS potential energy surface (PES) in Fig \ref{fig:dft} (b) by performing a 3D optimization for anionic OCS using B3LYP functional and 6-311++g(d,p) basis set. For OCS$^-$, the optimized bond angle is $135.5^{\circ}$ and C-O bond length is 1.75 \r{A} that is analogous to a C$_{2V}$ point group symmetry and agrees well with Gutsev \textit{et al.} \cite{gutsev1998el_eaffinity}. Optimisation indicates that the anionic GS is not achieved only through the parametrisation of the radial component. Practically, TNIs are formed in the electronically excited states where the normal coordinates attribute a rotation concerning the GS, resulting in axis-switching, and mode mixed Duschinsky effect \cite{zak2017ro_vibronic,sando2001ab,mebel1999ab,small1971herzberg} (mixed parallel and perpendicular transitions) that incorporates symmetrically perturbed vibrations and attributes interference pattern \cite{small1971herzberg,kim1999speed}. However, this allows rovibronic transitions in an electronically excited state, forbidden by usual rotational selection rules \cite{zak2017ro_vibronic} that leads to departure from the usual expressions of rotational line intensities and should be a rationale for observing the sudden intensity boost. 

\begin{figure}[ht!]
    \centering
    \includegraphics[scale=1.05]{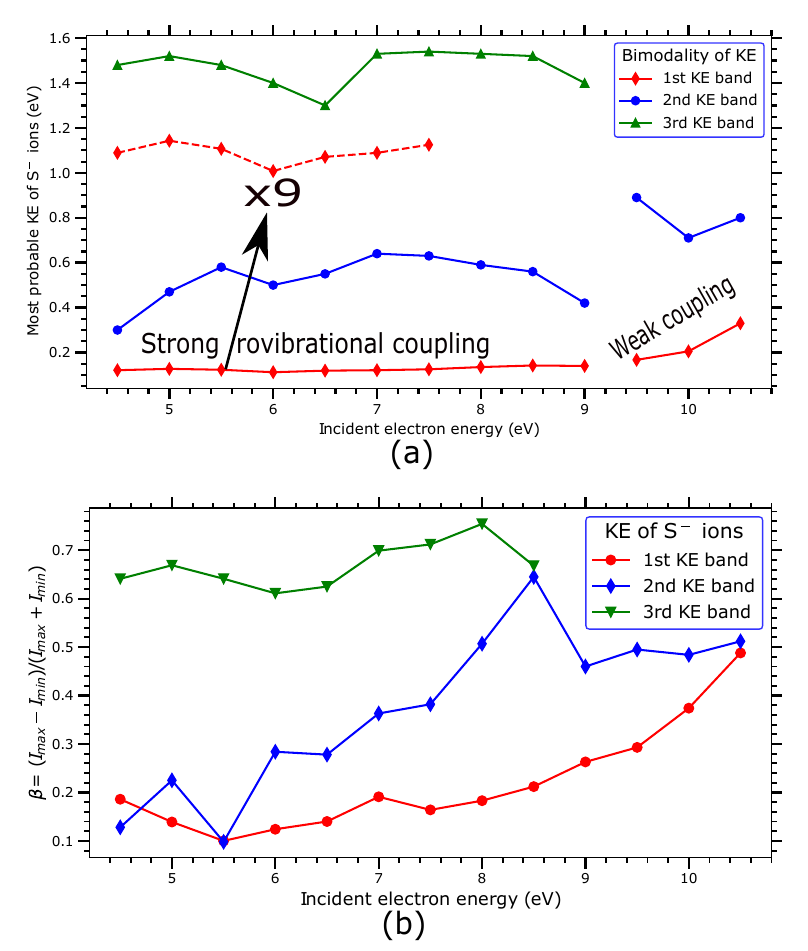}
    \caption{(a) Beam-energetic evolution for the most probable KE of S$^-$ fragments is shown that directly map (b) anitsotropy magnitude ($\beta$) of the ADs for different KE bands.}
    \label{fig:beta_aniso_pic}
\end{figure}

Thus, the KE energy bands of S$^-$ fragments are rovibrationally coupled with an electronically ground state for the CO counter fragments and an electronically excited CO at 10.0 eV resonance. This strong coupling limits the kinematics of S$^-$ fragments to be evolved more concerning the incident beam energies. It will prevent determining the band's threshold energy, mainly depending on the slope of the most probable kinetic energy positions of S$^-$anions. Interestingly, during the estimation of band's thresholds, we have observed a prominent bi-modal signature for each of the KE bands in the most probable KE distribution of S$^-$/OCS as depicted in Fig. \ref{fig:beta_aniso_pic}(a), matches well with other reports \cite{sato1995ion,suzuki1998nonadiabatic}. This bi-modality directly depends on the rotational state population for the fragmented CO at different incident electron energy that forcibly incorporates complexities in the rovibronic spectrum, mainly governed by axial rotation ($\omega$) followed by bending vibrations.

However, the rovibronic coupling would allow the dynamics to assign anisotropy in angular distributions (AD), which ingeniously depends on the electron-perturber interaction nature and is extracted using the same wedge slicing technique. Using transitional expressions for C$_{2V}$ point group symmetry as discussed by Ram \textit{et al.} \cite{ram2010thesis}, we have fitted the ADs of S$^-/$OCS. A1, A2, B1 \& B2 Operational symmetries are involved in the DEA dynamics for S$^-$/OCS, where A1 and B2 contribute significantly to the ADs for the first and second bands, A1 for the third band. We have employed sixth-order cosine and sine functions in the C$_S$ group fit expression \cite{ch3cho_dea_cs} that reveals A$^{\prime}$ contributes more than A$^{\prime \prime}$ where A$^{\prime \prime}$ is strongly associated at 6.0 eV resonance. It implies high \& low momentum sulfur fragments are associated with dissociation from the A$^{\prime}$ symmetric state, while dissociation from the A$^{\prime \prime}$ symmetric state only provides high-speed fragments, sounds good with Suzuki \textit{et al.} \cite{suzuki1998nonadiabatic}. Since the dissociation dynamics is a two-body (CO \& S$^-$) fragmentation, we have also determined the ADs in the framework of C$_{\infty V}$ point group symmetries and used the expressions from O'Malley \& Taylor \cite{omelly_1968} for modelling. The dominance of $\Sigma$ with $\Delta$ type resonances was observed in low energy bands over $\Sigma$ with $\Pi$ up to 9.0 eV. Fit alter for the high momentum KE bands that strongly attributes “Renner-Teller” vibronic splitting \cite{peric_rt_effect,brown_rt_hamiltonian,herzberg1933teller,renner1934theorie,herzberg2013molecular_rt,bunker2006molecular_rt,Herzberg_three,1966msms.book.....H} associated with $\Pi$ and $\Delta$ symmetries. Generally, the Electronic GS of a linear molecule associated with $\Sigma$ symmetry exhibits no coupling between electronic motion and vibration, whereas $\Pi$ exhibits reasonable coupling with the $\Delta$ symmetric states. 

Thus, the DEA dynamics of the OCS consisting of three fundamental building blocks can be studied using three different point group symmetries. We speculate that the symmetry reshuffles while using the mass-energy balance. The point group C$_{\infty V}$ switches to C$_S$ while considering the mass-difference between sulfur and oxygen atom. Both point groups directly correlate with the C$_{2V}$ point group symmetry during the formation of the TNI, which tests well with the report of Carter \textit{et al.} \cite{carter1998molecular} and is schematically represented in Fig \ref{fig:dft} (d), conventionalised as the partial correlation among point group symmetries. The angular anisotropy magnitude $\beta = \frac{I_{max}-I_{min}}{I_{max}+I_{min}}$ is unexpectedly proportional to the fragment's bi-modality in the most probable KE distribution for each of the momentum band as reflected in Fig. \ref{fig:beta_aniso_pic} (a) \& (b), and agrees well with the speed-dependent orientation effect modelled by Zare \textit{et al.} \cite{kim1999speed} that attributes quantum mechanical interference associated with in-plane mixed parallel and out-of-plane abrupt transitions. Unsurprisingly, two high momentum bands merge after 8.0 eV incident electron energy as reflected in VSIs, KE \& angular anisotropy distribution. It insists on assigning as CO is fragmented in its electronically excited state at 10.0 eV resonance \cite{iga1995,krupenie1966nat}, a rationale for observing high momentum tri-modal to low momentum bi-modal marginal effect for the S$^-$'s KE, and rovibronic coupling is less at this resonance that may allow us to determine band threshold. Crucially, it is non-trivial to report the dynamical time scale rearrangement and the axial recoil breakdown that reveals proper axial rotation with bending dissociation. Though, the axial recoil approximation holds fine when diatomic molecular ions are formed in a dissociative state \cite{wood1997limitations}, and falls short if the TNI possesses highly excited bending vibrations with appreciable rotational effects \cite{adachi1997renner}. So, TNI should possess beam-energy dependent unconventional axial rotation ($\omega$), which could be a rationale for observing similarity between KE and angular anisotropy in the same frame. Levine \textit{et al.} \cite{remacle2006electronic} nicely pointed out that during a chemical rearrangement of the bonded atoms, the motion of the nuclei sets the rearrangement time scale because fast electronic reorganisation exactly follows the shifts in the position of the slow nuclei. So, TNI's bent geometry through axial rotation followed by bond dissociation can reinforce this chemical rearrangement model. Thus, the rotational motion regulates the time scale for the dissociation of TNI.

\begin{table}[ht]
 \centering
\caption{Calculated vertical transitional resonance positions with the corresponding symmetries and excitation nature are tabulated.}
\label{tab:r_matrix}
\begin{tabular}{ccc}

\hline
\hline
    \textbf{\vtop{\hbox{\strut \ \ \ Vertical resonance }\hbox{\strut \hspace{8mm}position (eV) }}\hspace{3mm}}  & \textbf{\vtop{\hbox{\strut Symmetry}\hbox{\strut \ \ \ \ \ }}\hspace{3mm}}  & \textbf{\vtop{\hbox{\strut Exciatation}\hbox{\strut \hspace{3mm}nature }}\hspace{3mm}}\\
\hline
0.0 eV & A1  & Singlet  \\
5.54 eV & A1 & Triplet  \\

5.97 eV & A1+A2 & Triplet  \\ 

6.36 eV  & A2 & Triplet  \\

6.38 eV  & A2 & Singlet \\ 
6.43 eV  & A1+A2  & Singlet \\ 
8.72 eV  & B1+B2 & Triplet \\
9.47 eV  & B1+B2 & Singlet \\ 
9.98 eV  & A1 &  Triplet\\
\hline
\hline
\end{tabular}
\end{table}

Utilizing Quantemol$-$N \cite{PhysRevA.93.012702_minaxi,yadav2020low,tennyson2007quantemol}, we have applied space division based R-matrix method \cite{tennyson1984reson,tennyson2010electron} to determine the nature of DEA resonances for S$^-$/OCS. The vertical transitional resonance positions so obtained are listed in Table \ref{tab:r_matrix}. Notably, low-energy resonances are dominantly associated with A1 \& A2 while  B1 \& B2 associate with high-energy resonances, agreeing with our analysis.\\

\textit{Conclusions-} DEA dynamics for the vibration of fragmented CO attached to S$^-$/OCS, therefore, provides a fascinating opportunity to study rovibrational coupling in the context of the RTE and mixed-mode Duschinsky effect. Our ADs analysis associated with such linear and bent symmetries \cite{mcglynn1971electronic} strongly demands indirect observation of RTE. The ADs signify that the three constituting atoms with different masses behave like three-point groups and strongly support the substitution of an atom with a point. In future, DEA dynamics in other polyatomic molecules might show other engrossing features like bond stretch isomerism effect  \cite{homray2019bond_stretch} and importantly conical intersection mechanism \cite{longuet1975intersection,ruf2012high,worner2011conical,jost2002x2a1,slaughter2016_review,sanrey2006quantum}, whereas a consequence, the Born-Oppenheimer or adiabatic approximation \cite{handy1996adiabatic,born1968m,kolsos1970adiabatic} breaks down that results from the dynamics are dominated by a strong coupling between nuclear velocity and derivative non-adiabatic coupling vector field \cite{malhado2014non}.\\

\textit{Acknowledgement-} N. K. gratefully acknowledges the financial support from `DST of India' for the "INSPIRE Fellowship" program. We acknowledge Dr Kousik Samanta, Mr Subhasish Das, \& Abhisek Ghosal for their scientific discussion that guides us to reach the goal. We acknowledge financial support from the Science and Engineering Research Board (SERB) for supporting this research under  Project No. "CRG/2019/000872".\\

\bibliography{nk_bib} 
\onecolumngrid
\vspace{60mm}
\newpage
\section*{Supplementary}
\section{KE \& AD fit}
Here, we have shown the Gaussian fit KE distribution and the most probable peak energy values are used to determine the rovibrationally coupled bi-modality signature.
\begin{figure*}[hb!]
    \centering
    \includegraphics[scale=0.8]{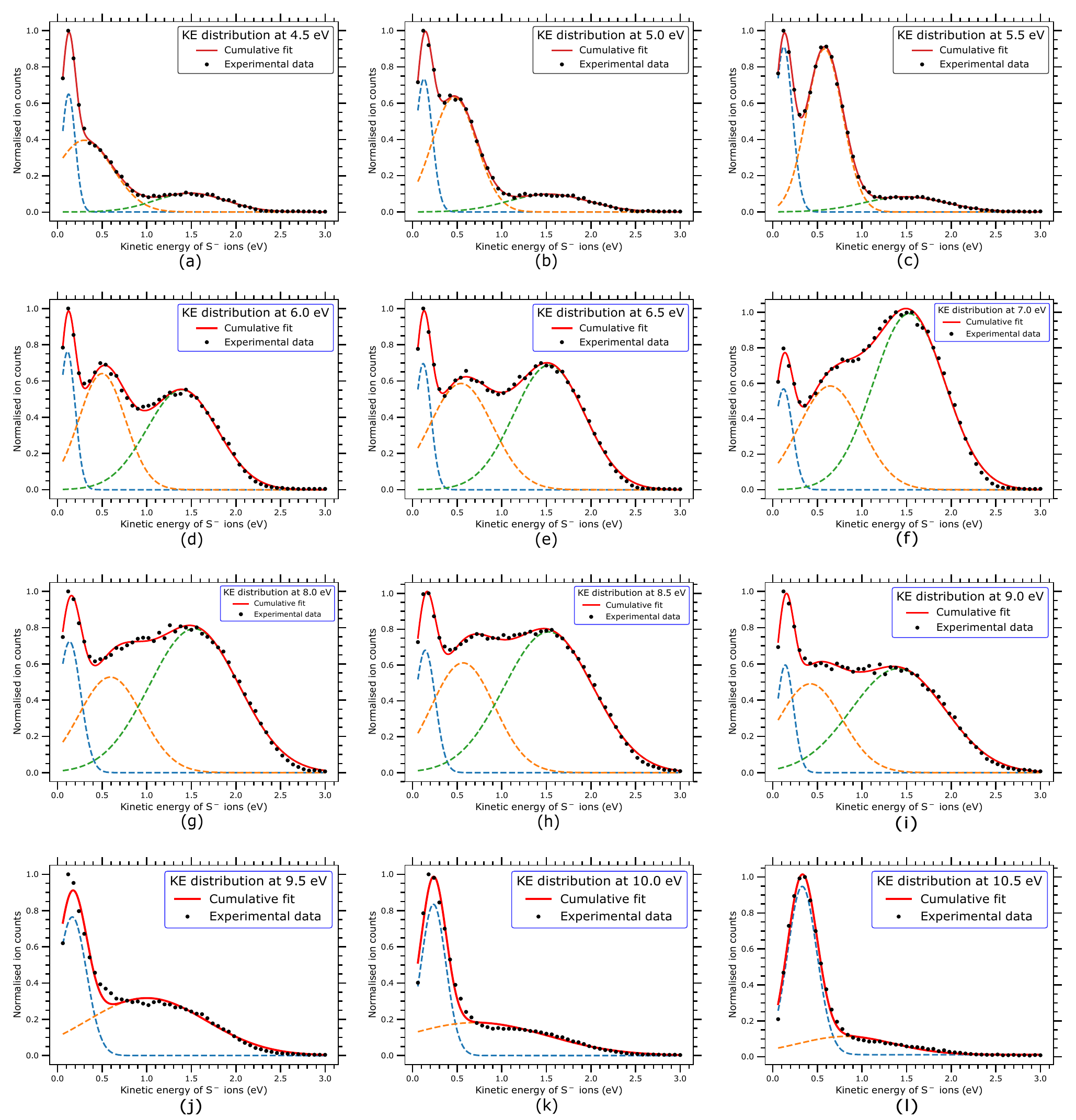}
    \caption{Figure reflects three Gaussian fit up-to 9.0 eV incident beam energy and two Gaussian for rest thrice. All the cumulative fit matches well with the experimental data.    }
    \label{fig:ke_fit}
\end{figure*}
\begin{figure*}
    \centering
    \includegraphics[scale=0.8]{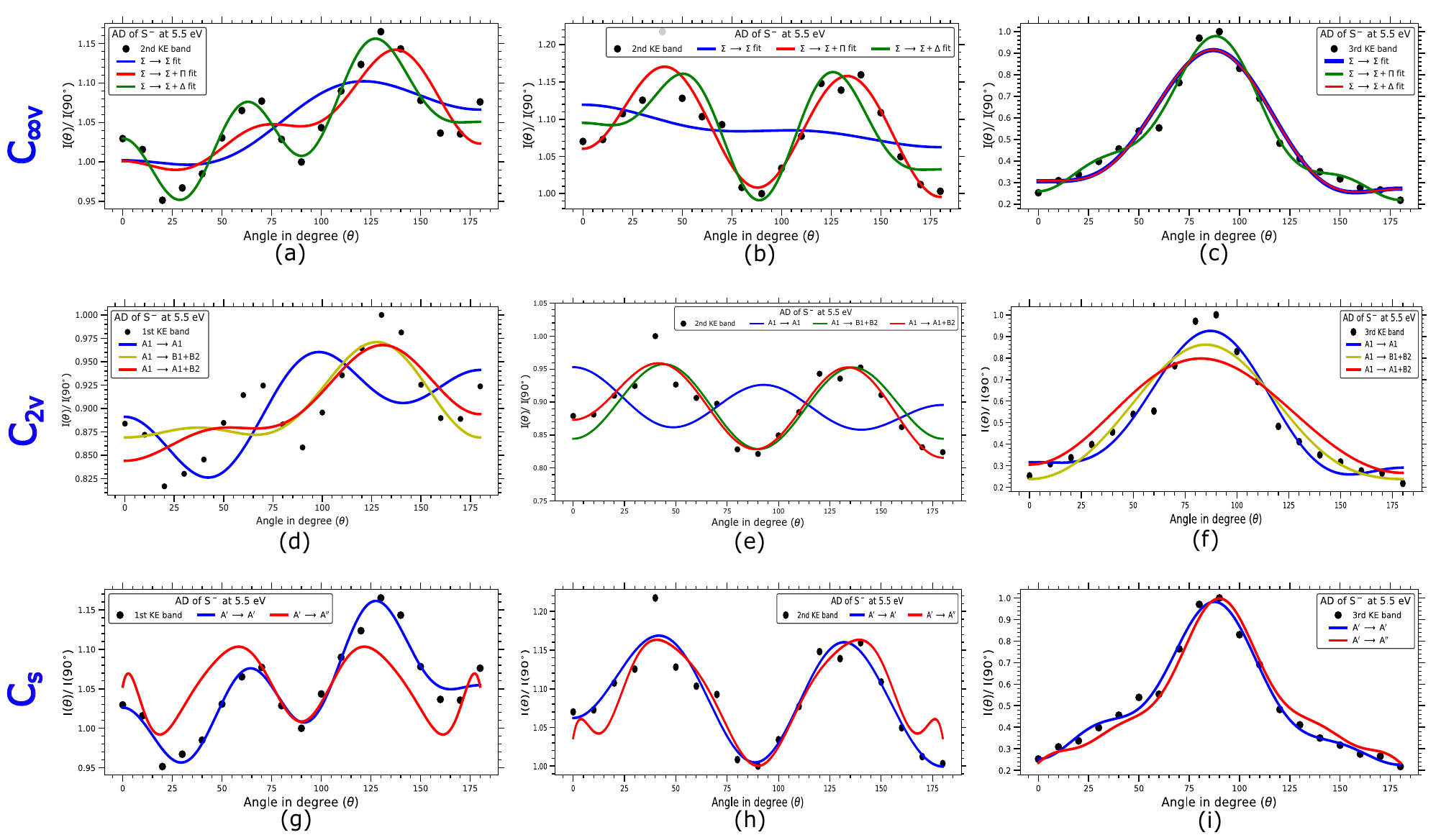}
    \caption{Figure represents the ADs of different KE bands for the nascent S$^-$ anionic figments at 5.5 eV incident electron energy as pointed out in the legend. The fits reveal final transitional states is dominantly associated with $\Sigma$ over $\Delta$ and $\Pi$ symmetries wherein $\Pi$ contributes more to the high momentum bands for C$_{\infty v}$ point group, A$^{\prime}$ over A$^{\prime\prime}$ wherein A$^{\prime\prime}$ contributes more to the high momentum bands for C$_{s}$ point group, and A1 \& B2 over A2 \& B1 but A1 contributes more to the high energetic bands for C$_{2v}$ point group.}
    \label{fig:fit_55eV}
\end{figure*}

\begin{figure*}
    \centering
    \includegraphics[scale=0.8]{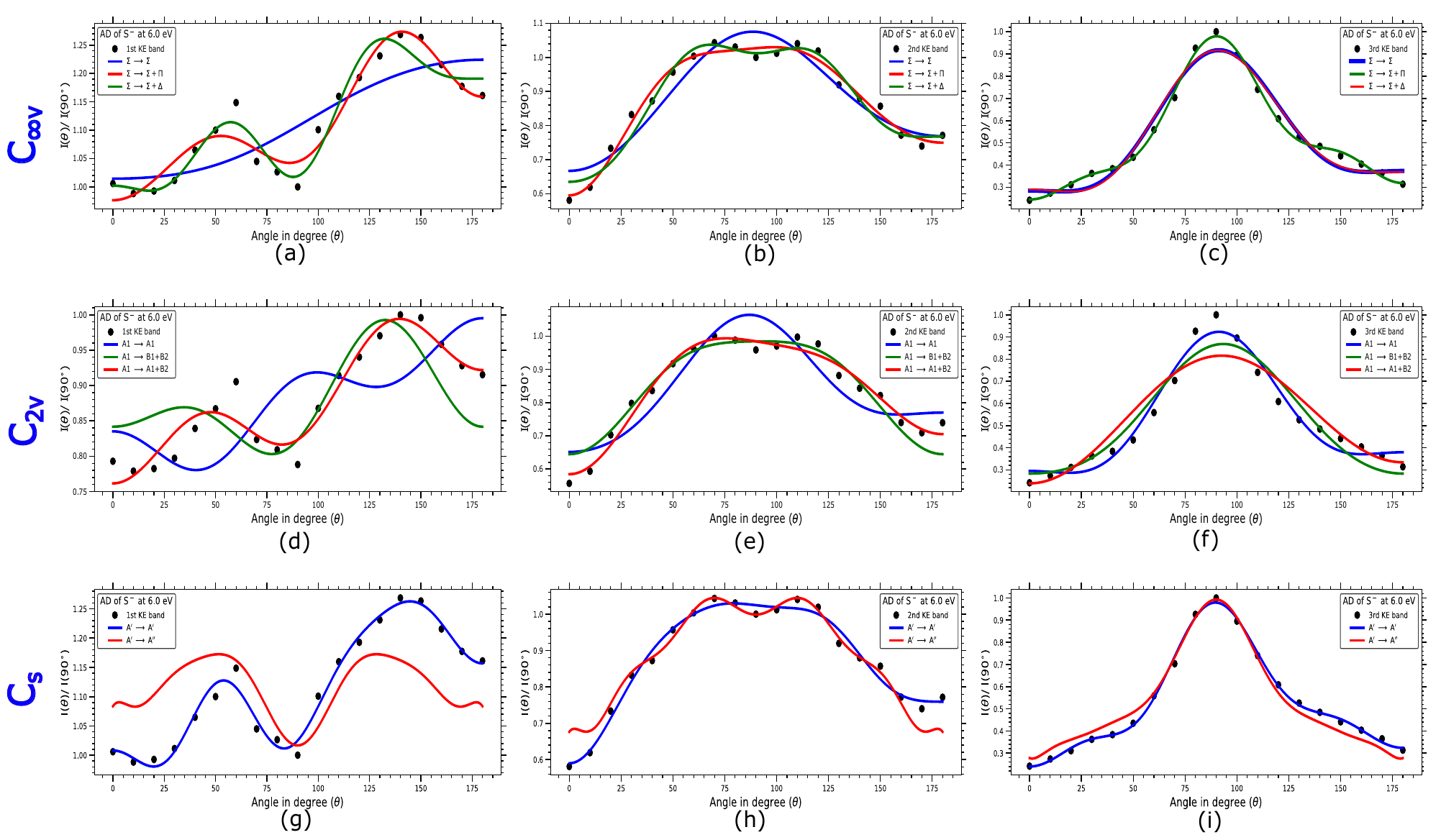}
    \caption{Figure represents the ADs of different KE bands for the nascent S$^-$ anionic figments at 6.0 eV incident electron energy as indicated in the legend. The fits reveal final transitional states are dominantly associated with $\Sigma$ over $\Delta$ and $\Pi$ symmetries wherein $\Pi$ contributes more to the high momentum bands for C$_{\infty v}$ point group, A$^{\prime}$ over A$^{\prime\prime}$ wherein A$^{\prime\prime}$ contributes more to the high momentum bands for C$_{s}$ point group, and A1 \& B2 over A2 \& B1 but A1 contributes more to the high energetic bands for C$_{2v}$ point group.}
    \label{fig:fit_6eV}
\end{figure*}

\begin{figure*}
    \centering
    \includegraphics[scale=0.8]{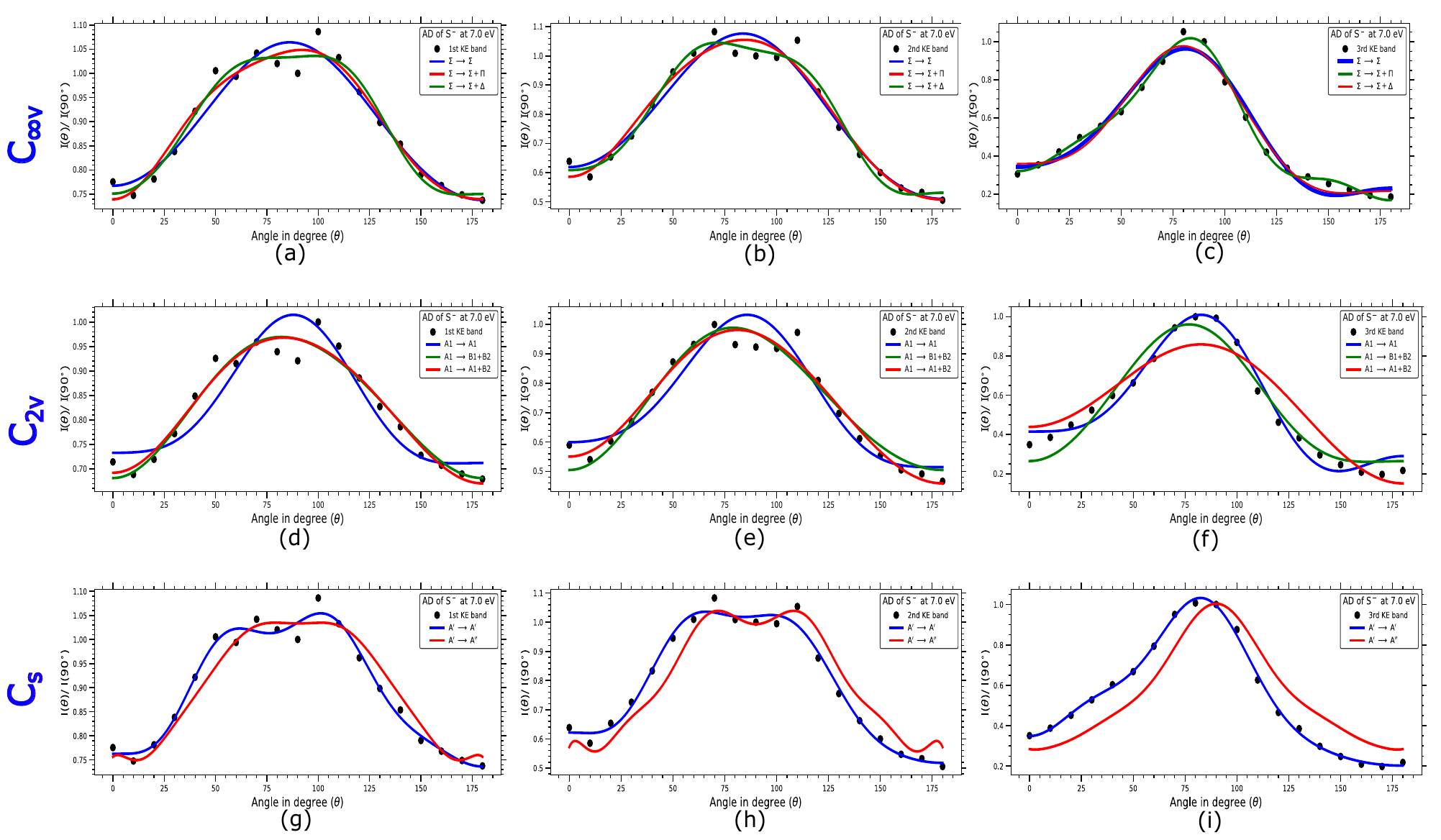}
    \caption{Figure represents the ADs of different KE bands for the nascent S$^-$ anionic figments at 7.0 eV incident electron energy as indicated in the legend box. The fits reveal final transitional states are dominantly associated with $\Sigma$ over $\Delta$ and $\Pi$ symmetries wherein $\Pi$ contributes more to the high momentum bands for C$_{\infty v}$ point group, A$^{\prime}$ over A$^{\prime\prime}$ wherein A$^{\prime\prime}$ contributes more to the high momentum bands for C$_{s}$ point group, and A1 \& B2 over A2 \& B1 but A1 contributes more to the high energetic bands for C$_{2v}$ point group.}
    \label{fig:fit_7eV}
\end{figure*}

\begin{figure*}
    \centering
    \includegraphics[scale=0.8]{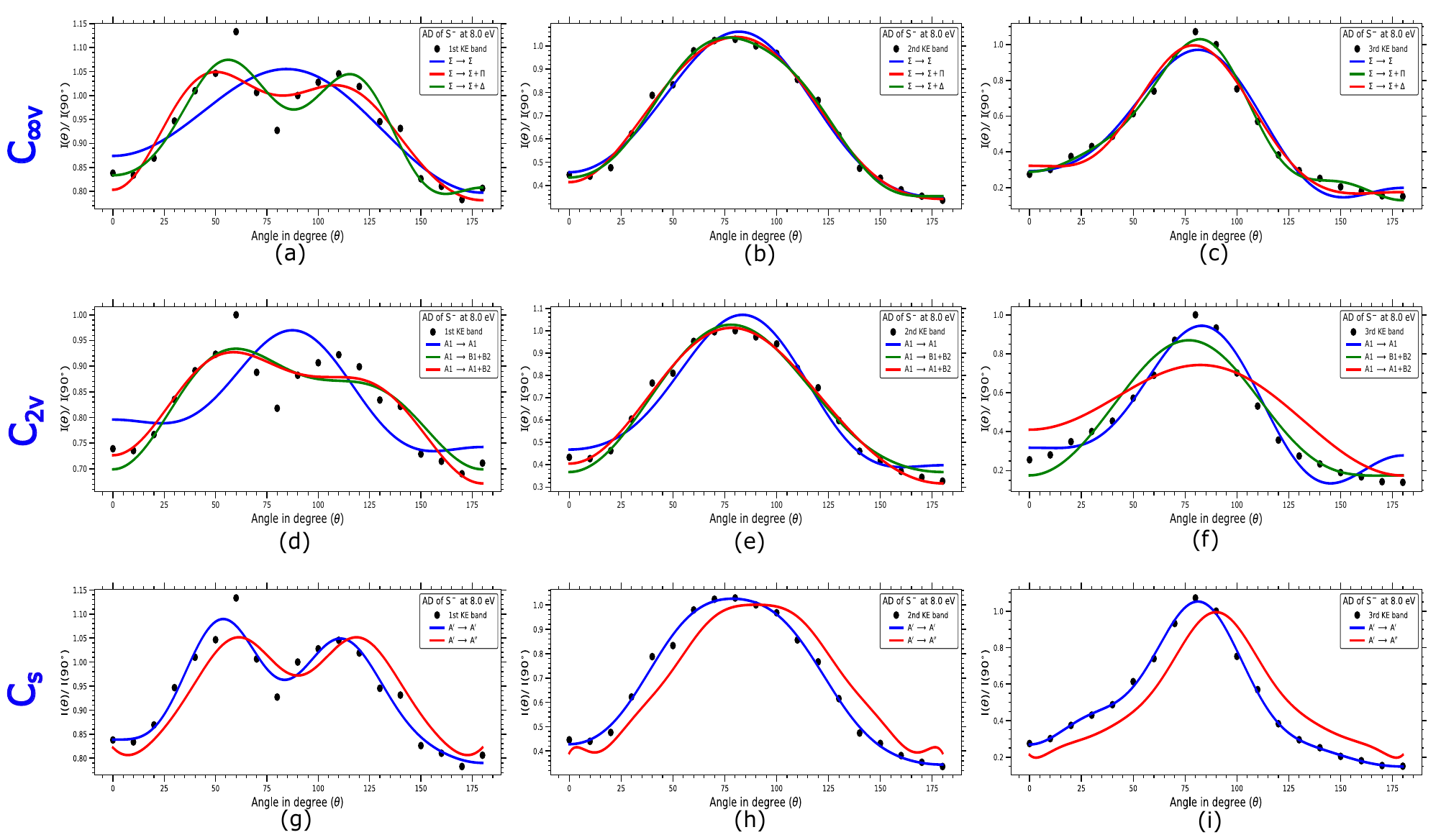}
    \caption{Figure represents the ADs of different KE bands for the nascent S$^-$ anionic figments at 8.0 eV incident electron energy as indicated in the legend. The fits reveal final transitional states are dominantly associated with $\Sigma$ over $\Delta$ and $\Pi$ symmetries wherein $\Pi$ contributes more to the high momentum bands for C$_{\infty v}$ point group, A$^{\prime}$ over A$^{\prime\prime}$ wherein A$^{\prime\prime}$ contributes more to the high momentum bands for C$_{s}$ point group, and A1 \& B2 over A2 \& B1 but A1 contributes more to the high energetic bands for C$_{2v}$ point group.}
    \label{fig:fit_8eV}
\end{figure*}

\begin{figure*}
    \centering
    \includegraphics[scale=0.8]{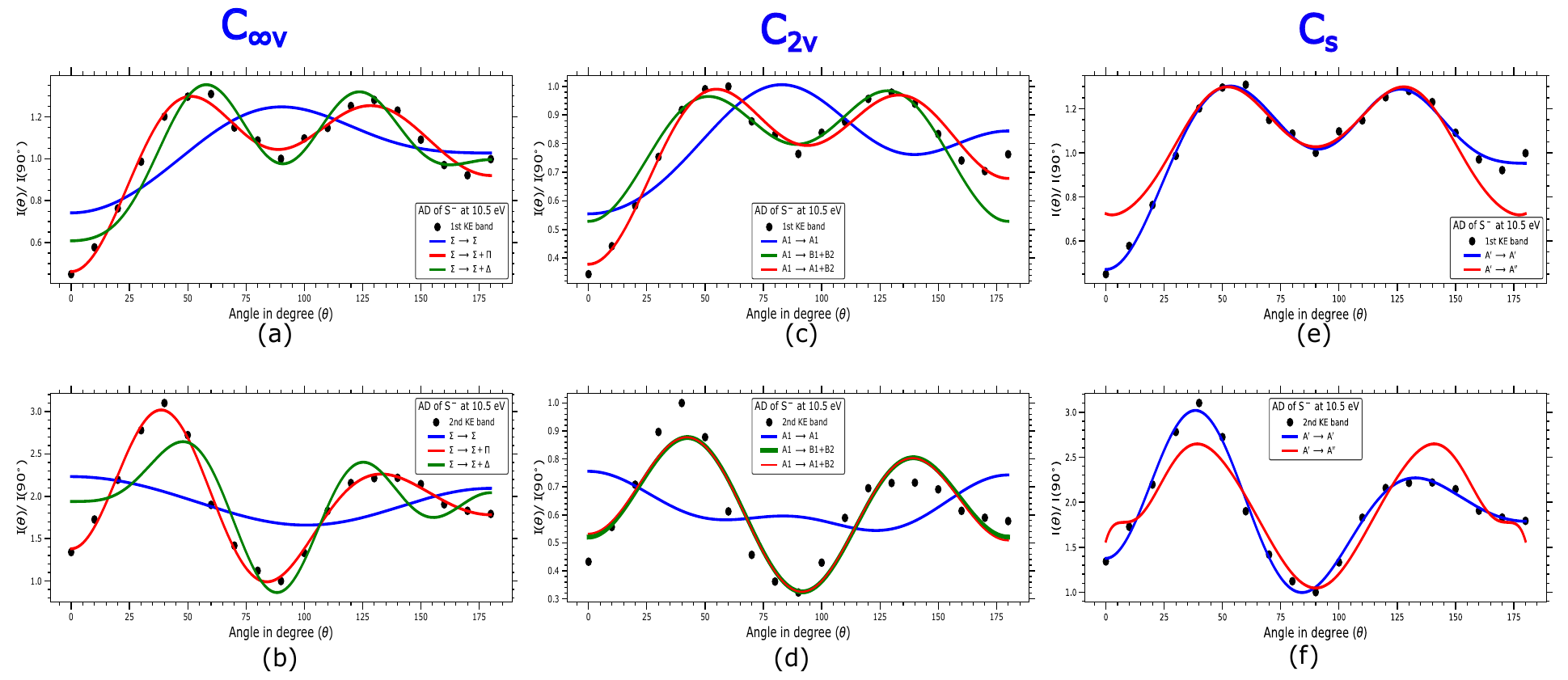}
    \caption{Figure represents the ADs of different KE bands for the nascent S$^-$ anionic figments at 10.5 eV incident electron energy as pointed out in the legend box. The fits reveal final transitional states are dominantly associated with $\Sigma + \Pi$ over $\Sigma + \Delta$ for C$_{\infty v}$ point group, A$^{\prime}$ over A$^{\prime\prime}$ for C$_{s}$ point group, and A1+B2 over B1+B2 for C$_{2v}$ point group. The absence of high momentum band at 10.5 resonance structure strongly implies that highly energetic S$^-$ anions are mostly associated with A1 symmetry for C$_{2v}$ point group and A$^{\prime\prime}$ symmetry for C$_s$ point group.}
    \label{fig:fit_10_5eV}
\end{figure*}

\begin{table}[!ht]
    \centering
    \caption{Calculated fit parameter are tabulated here for different KE bands as a function of electron beam energy using three different point group symmetries.}
    \begin{tabular}{|c|c|c|c|c|c|}
    \hline
        ~ Electron energy & Point group & Kinetic energy Bands & Best fit symmetries & $\hspace{3mm}$ R$^2$ value $\hspace{3mm}$ & Adj. R$^2$ value \\ \hline
        \multirow{9}{*}{5.5 eV} & \multirow{3}{*}{C$_{2v}$} & 1st band & A1 $\to$ A1+B2 & 0.641 & 0.503 \\ \cline{3-6}
        ~ &  & 2nd band & A1 $\to$ A1+B2 & 0.903 & 0.866 \\ \cline{3-6}
        ~ &  & 3rd band & A1 $\to$ A1 & 0.951 & 0.932 \\ \cline{2-6}
        ~ & \multirow{3}{*}{C$\infty v$} & 1st band & $\Sigma \to \Sigma + \Delta$ & 0.946 & 0.860 \\ \cline{3-6}
        ~ &  & 2nd band & $\Sigma \to \Sigma + \Pi$ & 0.905 & 0.755 \\ \cline{3-6}
        ~ &  & 3rd band & $\Sigma \to \Sigma + \Pi$ & 0.988 & 0.970 \\ \cline{2-6}
        ~ & \multirow{3}{*}{C$_s$} & 1st band & A$^{\prime}$ $\to$ A$^{\prime}$ & 0.950 & 0.871 \\ \cline{3-6}
        ~ &  & 2nd band & A$^{\prime}$ $\to$ A$^{\prime}$ & 0.907 & 0.761 \\ \cline{3-6}
        ~ &  & 3rd band & A$^{\prime}$ $\to$ A$^{\prime}$ & 0.989 & 0.972 \\ \hline
        \multirow{9}{*}{6.0 eV} & \multirow{3}{*}{C$_{2v}$} & 1st band & A1 $\to$ A1+B2 & 0.919 & 0.889 \\ \cline{3-6}
        ~ &  & 2nd band & A1 $\to$ A1+B2 & 0.970 & 0.959 \\ \cline{3-6}
        ~ &  & 3rd band & A1 $\to$ A1 & 0.957 & 0.941 \\ \cline{2-6}
        ~ & \multirow{3}{*}{C$\infty v$} & 1st band & $\Sigma \to \Sigma + \Delta$ & 0.939 & 0.843 \\ \cline{3-6}
        ~ &  & 2nd band & $\Sigma \to \Sigma + \Pi$ & 0.978 & 0.945 \\ \cline{3-6}
        ~ &  & 3rd band & $\Sigma \to \Sigma + \Pi$ & 0.996 & 0.990 \\ \cline{2-6}
        ~ & \multirow{3}{*}{C$_s$} & 1st band & A$^{\prime}$ $\to$ A$^{\prime}$ & 0.979 & 0.945 \\ \cline{3-6}
        ~ &  & 2nd band & A$^{\prime}$ $\to$ A$^{\prime}$ & 0.981 & 0.951 \\ \cline{3-6}
        ~ &  & 3rd band & A$^{\prime}$ $\to$ A$^{\prime}$ & 0.997 & 0.992 \\ \hline
        \multirow{9}{*}{7.0 eV} & \multirow{3}{*}{C$_{2v}$} & 1st band & A1 $\to$ A1+B2 & 0.955 & 0.938 \\ \cline{3-6}
        ~ &  & 2nd band & A1 $\to$ A1+B2 & 0.965 & 0.952 \\ \cline{3-6}
        ~ &  & 3rd band & A1 $\to$ A1 & 0.972 & 0.961 \\ \cline{2-6}
        ~ & \multirow{3}{*}{C$\infty v$} & 1st band & $\Sigma \to \Sigma$ & 0.946 & 0.926 \\ \cline{3-6}
        ~ &  & 2nd band & $\Sigma \to \Sigma + \Delta$ & 0.979 & 0.947 \\ \cline{3-6}
        ~ &  & 3rd band & $\Sigma \to \Sigma + \Pi$ & 0.995 & 0.988 \\ \cline{2-6}
        ~ & \multirow{3}{*}{C$_s$} & 1st band & A$^{\prime}$ $\to$ A$^{\prime}$ & 0.981 & 0.952 \\ \cline{3-6}
        ~ &  & 2nd band & A$^{\prime}$ $\to$ A$^{\prime}$ & 0.984 & 0.958 \\ \cline{3-6}
        ~ &  & 3rd band & A$^{\prime}$ $\to$ A$^{\prime}$ & 0.997 & 0.993 \\ \hline
        \multirow{9}{*}{8.0 eV} & \multirow{3}{*}{C$_{2v}$} & 1st band & A1 $\to$ A1+B2 & 0.853 & 0.796 \\ \cline{3-6}
         &  & 2nd band & A1 $\to$ A1+B2 & 0.991& 0.988 \\ \cline{3-6}
         &  & 3rd band & A1 $\to$ A1 & 0.946 & 0.925 \\ \cline{2-6}
         & \multirow{3}{*}{C$\infty v$} & 1st band & $\Sigma \to \Sigma + \Delta$ & 0.925 & 0.808 \\ \cline{3-6}
         &  & 2nd band & $\Sigma \to \Sigma + \Delta$ & 0.993 & 0.982 \\ \cline{3-6}
         &  & 3rd band & $\Sigma \to \Sigma + \Pi$ & 0.994 & 0.984 \\ \cline{2-6}
         & \multirow{3}{*}{C$_s$} & 1st band & A$^{\prime}$ $\to$ A$^{\prime}$ & 0.945 & 0.859 \\ \cline{3-6}
         &  & 2nd band & A$^{\prime}$ $\to$ A$^{\prime}$ & 0.994 & 0.985 \\ \cline{3-6}
         &  & 3rd band & A$^{\prime}$ $\to$ A$^{\prime}$ & 0.998 & 0.994 \\ \hline
         \multirow{6}{*}{10.5 eV} & \multirow{2}{*}{C$_{2v}$} & 1st band & A1 $\to$ A1+B2 & 0.964 & 0.949 \\ \cline{3-6}
         &  & 2nd band & A1 $\to$ B1+B2 & 0.836 & 0.775 \\ \cline{2-6}
         & \multirow{2}{*}{C$\infty v$} & 1st band & $\Sigma \to \Sigma + \Pi$ & 0.979 & 0.947 \\ \cline{3-6}
         &  & 2nd band & $\Sigma \to \Sigma + \Pi$ & 0.988 & 0.969 \\ \cline{2-6}
         & \multirow{2}{*}{C$_s$} & 1st band & A$^{\prime}$ $\to$ A$^{\prime}$ & 0.988 & 0.970 \\ \cline{3-6}
         &  & 2nd band & A$^{\prime}$ $\to$ A$^{\prime}$ & 0.988 & 0.969 \\ \hline

    \end{tabular}
\end{table}

\end{document}